\documentclass[onecolumn]{aastex631}
\usepackage{graphicx}
\usepackage{psfrag}
\usepackage{amsmath}
\usepackage{amssymb}
\usepackage{color,xcolor}
\usepackage{hyperref}
\hypersetup{
    colorlinks=true,
    citecolor=blue,
    filecolor=black,
    linkcolor=blue,
    urlcolor=magenta,
    linktocpage=true,
    breaklinks = true
}

\usepackage{enumitem}
\usepackage{subfigure,float}
\usepackage{multirow}
\usepackage{makecell}
\usepackage[normalem]{ulem}

\usepackage{array,booktabs}
\usepackage{mathtools}

\def \de {\mathrm{DE}}

\def \o {\mathcal{O} }

\newcommand{\beq}{\begin{equation}}
\newcommand{\eeq}{\end{equation}}

\newcommand{\ber}{\begin{eqnarray}}
\newcommand{\eer}{\end{eqnarray}}

\newcommand{\ba}{\begin{align}}
\newcommand{\ea}{\end{align}}
\newcommand{\wprobetitle}{\textup{$w_0$-probe}}

\def \lcdm {$\Lambda$CDM~}

\def \omz {Om(z)}

\def \wpt {$w_0$-probe~}

\def\w0{w_0}
\def \w {w}
\def\om0{\Omega_{{0\rm m}}}
\def \omegr0{\Omega_{{0\rm r}}}

\def \wzp {w_0\text{-}{\rm probe}} 

\begin{document}
\title{$w_0$-probe: A new diagnostic of dark energy based on $Om$} 

\author[0000-0003-0141-606X]{Satadru Bag}
\email{satadru.bag@tum.de}
\affiliation{Technical University of Munich, Physics Department,  James-Franck-Stra{\ss}e 1, 85748 Garching, Germany}
\affiliation{Max-Planck-Institut f{\"u}r Astrophysik, Karl-Schwarzschild Stra{\ss}e 1, 85748 Garching, Germany}

\author[0000-0002-0862-8789]{Ryan E. Keeley}
\affiliation{Department of Physics and Astronomy, University of California, Irvine, CA 92697-4575, USA}

\author[0000-0002-3862-4244]{Varun Sahni}
\affiliation{Inter-University Centre for Astronomy and Astrophysics, Post Bag 4, Ganeshkhind, Pune
411 007, India}

\author[0000-0001-6815-0337]{Arman Shafieloo}
\affiliation{Korea Astronomy and Space Science Institute, Daejeon 34055, Korea}
\affiliation{University of Science and Technology, Daejeon 34113, Korea}

\begin{abstract}
Recent DESI data suggest that dark energy may be evolving and motivate the use of model-independent diagnostics such as $Om(z)$ and probes of the equation of state (EoS) of dark energy, $w(z)$. Traditional reconstructions of $w(z)$ rely on differentiating the expansion history, $h(z)=H(z)/H_0$, which amplifies noise and systematic uncertainties. In this work, we introduce a new diagnostic, the $w_0$-probe, which is constructed from $Om(z)$, and  which enables a direct determination of the current EoS from $h(z)$ without any additional differentiation. While retaining the null-test capability of $Om(z)$ for $\Lambda$CDM, the $w_0$-probe also provides a direct estimate of $w_0$ -- the current EoS of dark energy. We demonstrate that this reconstruction of $w_0$ is robust for any smooth underlying $w(z)$. We apply this method to Gaussian-process (GP) reconstructions of $h(z)$ using current SNe Ia+BAO+CMB data. Both $\omz$ and the \wpt exclude \lcdm at a Gaussian-equivalent significance of
$\simeq3\sigma$, with the latter favoring $w_0 \simeq -0.62 \pm
0.21$ at the $95\%$ confidence level. To mitigate potential
over-constraining from GP priors, we additionally analyze $\chi^2$-limited
reconstructions with likelihoods exceeding the $95\%$ confidence threshold of the CPL fit. Among
these, more than $99.9\%$ of the realizations best supported by the data yield
$w_0>-1$ as $z\to0$, with the central $95\%$ lying in $w_0\in(-0.85,-0.40)$,
demonstrating the robustness of our results. The \wpt therefore provides a
simple, model-independent, and robust diagnostic of the current EoS of dark
energy.

\end{abstract}

\section{Introduction}

Understanding the nature of dark energy remains one of the central challenges in modern cosmology \citep{de1,varun_de1,paddy_de,Peebles_de,varun_de_dm}. Whether the accelerated expansion of the Universe is driven by a cosmological constant ($\Lambda$) or by a dynamical component continues to be an open question with far-reaching implications for fundamental physics. The possibility of evolving dark energy has received renewed attention in recent years. For instance, late-time phantom-like evolution has been discussed as a potential solution in alleviating the Hubble tension \citep{late-silk,late-ujjaini,late-melchiorri1,late-arman1,late-arman2,Li:2019yem,late-vagnozzi,late-valentino,Bag:2021cqm}. At the same time, several recent analyses \citep{desi_dr2_cosmology, Rodrigo2024,lodha2025} of the latest Dark Energy Spectroscopic Instrument (DESI) \citep{desi_dr1} data suggest a possible deviation from the standard spatially flat $\Lambda$ cold dark matter ($\Lambda$CDM, hereafter) cosmological model in the opposite direction, indicating non-phantom (or ``quintessence-like'') behavior of dark energy near the present epoch. These developments highlight the importance of model-independent probes that can test the properties of dark energy without relying on specific models or parametrizations.

In this context, the $Om(z)$ diagnostic provides a simple yet powerful
way to test departures from the standard \lcdm model \citep{om,om3,Omm}. Constructed purely from the expansion
history, it acts as a null test for the \lcdm model,
since in this case $Om(z)$ remains constant and equals $\om0$.
Any redshift dependence in $Om(z)$ therefore signals a deviation from
this standard model. Beyond serving as a null test, however, $Om(z)$ contains additional information about the underlying dark energy dynamics. In this work, we show that $Om(z)$ can be used to construct a new quantity, the $w_0$-probe, which enables a direct determination of the present-day equation of state (EoS) parameter, $w_0$, through (\ref{eq:w0probe}).

Traditionally, reconstructing the dark energy EoS requires differentiating the expansion history $h(z)\equiv H(z)/H_0$, typically inferred from discrete observational data such as SNe Ia, BAO, and CMB measurements. Differentiation amplifies noise and systematic uncertainties, rendering the reconstruction unstable, especially at low redshift where precision is most critical. The $w_0$-probe we introduce is built directly from $Om(z)$ and therefore depends only on $h(z)$ itself, without requiring additional differentiation. A primary goal of this work is to establish, on firm mathematical grounds, that for any smooth underlying $w(z)$, the $w_0$-probe evaluated at sufficiently low redshift converges to the true present-day EoS, $w_0$. We further examine higher-order corrections, quantifying the theoretical and systematic uncertainties when the probe is evaluated at moderate redshifts, $z \lesssim 0.5$.

We demonstrate the utility of this approach using Gaussian-process reconstructions of the normalized expansion history $h(z)$, based on up-to-date SNe Ia, BAO, and CMB data. Both GP prior-driven posterior samples and likelihood-selected realizations show deviations from $\Lambda$CDM when analyzed through $Om(z)$ and the $w_0$-probe. Looking ahead, forthcoming improvements in low-redshift measurements of the expansion history, including those anticipated from next-generation gravitational-wave standard siren observations \citep{2023PhRvD.108d4038S}, will allow even tighter reconstructions of $h(z)$. In such a scenario, the $w_0$-probe provides a simple and robust pathway toward increasingly precise determinations of the present-day properties of dark energy.

\section{Dark energy diagnostics}
\subsection{$Om(z)$ family of dark energy diagnostics }

\begin{figure}
 \centering
\includegraphics[width=0.5\linewidth]{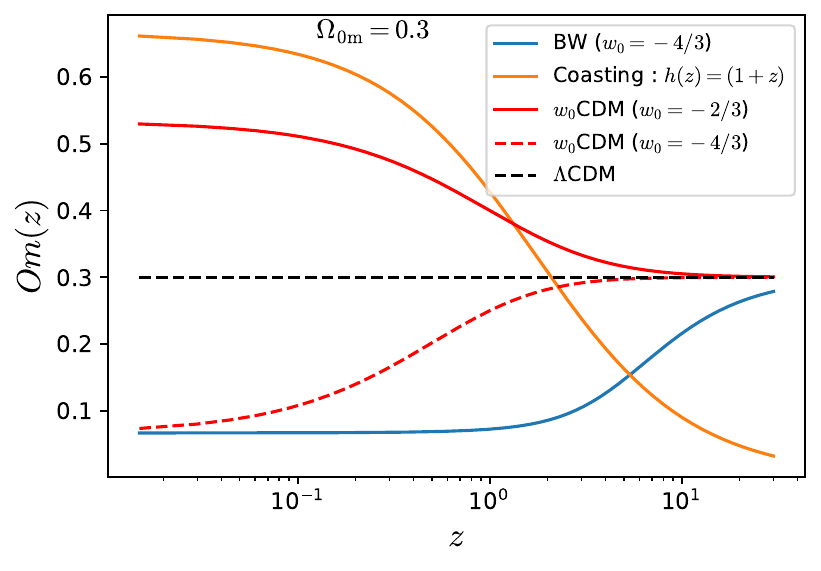}
\caption{$Om(z)$ for different dark energy models.}
\label{fig:omz_models}
\end{figure}

The $Om$ diagnostic is defined as \citep{om,Zunckel2008,om3,Omm}
\beq
Om(z) = \frac{h^2(z) - 1}{(1+z)^3 - 1}\;,
\label{eq:Om}
\eeq
where $h(z) = H(z)/H_0$. The $Om$ diagnostic has the interesting property that $Om(z)=\text{constant}=\om0$ only for the flat $\Lambda$ cold dark matter model (\lcdm for simplicity hereafter). For any evolving dark energy (DE) models one has $Om \neq \om0$. This can be used to directly test if \lcdm model is consistent with the expansion history reconstructed from a dataset such as the Pantheon+ \citep{PantheonP2} or Union3 \citep{Rubin:2023jdq} compilation of SNe~Ia.

The effectiveness of the $Om$ diagnostic in distinguishing between different dark energy models has been extensively discussed in the literature \citep{om, om3,Omm}. Before embarking on the main body of our paper we show how $Om$ can easily distinguish between several distinct dark energy models. Consider first a {\em coasting cosmology} with $a \propto t$ \citep{Sahni:1991ks}. This model has been suggested as an alternative to \lcdm in \citet{Melia_2011,Melia_2024}. It is easy to show that, since $H(z) = H_0 (1+z)$, the $Om$ diagnostic in coasting cosmology takes the simple form
\beq
Om(z) = \frac{(1+z)^2 - 1}{(1+z)^3 - 1}\;,
\label{eq:coast}
\eeq
which has the asymptotic limits
\beq
Om(z) =
\begin{cases}
\frac{2}{3}, & z=0 \\
0, & z \gg 1
\end{cases}
\eeq
By contrast, in conventional dark energy models
\beq \label{eq:om_matter_dom}
Om(z \gg 1) \simeq \om0\,,
\eeq
reflecting the presence of a matter dominated stage prior to cosmic acceleration.

In Fig.~\ref{fig:omz_models} we show the $Om(z)$ diagnostic for several dark energy models: (i) $\Lambda$CDM, (ii) $w_0$CDM with $w_0=-2/3$ and $w_0=-4/3$, (iii) phantom braneworld model with $w_0=-4/3$, and (iv) a coasting cosmology with $a \propto t$. Except for the rather peculiar coasting model, all other cases satisfy $Om(z)\to \om0$ at sufficiently large redshift ($z\gg1$). However, only in the case of \lcdm does $Om(z)$ remain equal to $\om0$ at all redshifts.

A clear trend emerges when comparing models with different equations of state. For quintessence-type dark energy with $w>-1$, $Om(z)$ approaches the \lcdm limit from above as $z$ increases, always remaining larger than $\om0$. By contrast, for models that are phantom-like ($w<-1$), $Om(z)$ approaches the same limit from below and remains smaller than $\om0$.

Therefore, $Om(z)$ not only provides a null test of the concordance \lcdm model, but its redshift evolution can also distinguish between different dark energy models, particularly between two fundamentally different classes of evolving dark energy: quintessence-type models with $w_0>-1$ and phantom-type models with $w_0<-1$.

Several related diagnostics can be constructed in a similar spirit to $Om(z)$ \citep{om,om3,Omm}. For instance, one may consider the difference or the ratio evaluated at two redshifts. It is easy to see that $Om(z_2)-Om(z_1) = 0$ while $Om(z_2)/Om(z_1) = 1$ in $\Lambda$CDM. One may also construct a two-point $Om$ diagnostic involving pairs of redshifts, and consider its differences or ratios. Such variants can enhance the sensitivity of the $Om$ approach to departures from \lcdm and help distinguish between different dark energy scenarios. In the next section we build upon this qualitative discussion and quantify the diagnostic power of $Om(z)$.
\subsection{$\omz$ as a direct probe of $w_0$}

For any generic smooth $w(z)$ the $w_0$-probe is defined as 
\begin{align} 
\wzp (z)\, \equiv \frac{Om(z) - 1}{1-\om0} &= \frac{h^2 - (1+z)^3}{(1-\om0)[(1+z)^3 -1]} \label{eq:w0probe}\\ 
&= w_0  + C_1 z + C_2 z^2 + \hdots\;,\label{eq:w0probe2}
\end{align}
where the coefficients $C_1, C_2, \hdots$ depend only on the exact shape of
$w(z)$, as we will see in Sec.~\ref{sec:derivation}.

Interestingly, for constant $w(z)=w_0$, there exist two fixed points, $w_0=-1$ (the $\Lambda$CDM model) and $w_0=0$, for which $Om(z)$ is also constant:
\begin{equation}\label{eq:wzp_fixed_pts}
w_0 \in \{0,-1\}
\;\Longrightarrow\;
Om(z) \in \{1,\om0\}
\;\Longrightarrow\; \wzp(z)=w_0 \quad \forall z .
\end{equation}
Substituting the values of $Om(z)$ into Eq.~\eqref{eq:w0probe}, one finds that $\wzp(z)$ exactly reduces to $w_0$ in these two limits, as indicated by the final implication above.

More generally, $(Om(z)-1)/(1-\om0)=w_0 + \o (z)$ which enables a model-independent probe of $w_0$ at $z \ll 1$, provided $\om0$ is constrained with reasonable accuracy. We therefore refer to this quantity as the $w_0$-probe. However, in the opposite limit, $z \gg 1$, corresponding to the matter-dominated era, Eq.~\eqref{eq:om_matter_dom} implies that $\wzp(z)$ asymptotically approaches $-1$,
\beq\label{eq:wzp_mat_dom}
\text{at } z \gg 1:\qquad Om(z)\to \om0 \;\Longrightarrow\; \wzp(z)\to -1 \;.
\eeq
Note that, this asymptotic behavior is generic to any cosmological model that undergoes a matter-dominated phase prior to the late-time accelerated expansion, and is not restricted to the $\Lambda$CDM.
The accuracy of $\wzp$ has been demonstrated in Appendix \ref{app:w0probe_extra} for a few dark energy models and parametrizations: $w_0$CDM, CPL, quintessence, PEDE, phantom braneworld.

\subsection{Detailed derivation}\label{sec:derivation}
Expressing the dark energy density as 
\begin{align}
  \Omega_{\de}(z) &= \Omega_{\de,0}\, f_{\de}(z),~\text{where}~ \Omega_{\de,0}=1-\om0,~\text{and}\, \label{eq:Ode1}\\
  ~f_{\de}(z)& \equiv \rho_{\rm DE}(z)/\rho_{{\de},0} =\exp\!\left[
3 \int_{0}^{z} \frac{1 + w(z')}{1+z'} \, dz'
\right] \label{eq:fde_def}
\end{align}
we obtain
\begin{align}
 Om(z)-1&=\frac{\om0 (1+z)^3 + (1-\om0)f_{\de}(z) -(1+z)^3 }{(1+z)^3  -1} \\
\implies \wzp(z)= \frac{Om(z)-1}{1-\om0}&= -1 +\frac{f_{\de}(z)-1}{(1+z)^3  -1} \label{eq:w_0_0}\;.
\end{align}
Equation~\eqref{eq:w_0_0} shows that the explicit dependence on
$\om0$ cancels identically, so that $\wzp(z)$ depends only on $f_{\de}(z)$ and
hence, through Eq.~\eqref{eq:fde_def}, only on $w(z)$. Consequently, $\wzp(z)$
is independent of $\om0$ whenever $w(z)$ is, as for phenomenological
parametrizations such as CPL \citep{CPL1,CPL2}. In physical models in which $w(z)$, as
determined by the dynamics, also depends on $\om0$, $\wzp(z)$ inherits this
dependence only implicitly, through $w(z)$.

We further note that, in model-independent analyses based solely on the
expansion history, such as the present one, dark energy cannot be
distinguished from other components without assuming a specific model.
Everything other than pressureless matter scaling as $(1+z)^3$ is therefore
necessarily treated as dark energy, with an effective EoS
\begin{equation}\label{eq:weff}
    w(z) \equiv \frac{p_{\rm DE}}{\rho_{\rm DE}}=\frac{2(1+z)h'(z)/h(z)-3}{3\left[1-\om0(1+z)^3/h^2(z)\right]} \;,
\end{equation}
where $h'(z)\equiv dh(z)/dz$. 

Any additional contribution to the expansion
rate, such as spatial curvature or interactions in the dark sector, is thus
effectively absorbed into $f_{\de}(z)$ in Eq.~\eqref{eq:Ode1}, with
$\Omega_{\de,0}=1-\om0$, and Eq.~\eqref{eq:w_0_0} holds exactly.
Throughout this work, $w(z)$ and $w_0$ therefore refer to this effective
EoS.

Let us evaluate the right hand side (RHS) of Equation~\eqref{eq:w_0_0} for a generic $w(z)$. Any continuous and smooth $w(z)$ can be Taylor expanded around $z=0$,
\begin{equation}
w(z) = w_0 + \left.\frac{dw}{dz}\right|_{z=0} z + \frac{1}{2} \left.\frac{d^2w}{dz^2}\right|_{z=0} z^2 + \cdots =\sum_{n=0}^{\infty} \frac{w_n}{n!}\, z^n \;,~\text{where}~w_n \equiv \left.\frac{d^n w}{dz^n}\right|_{z=0}.
\end{equation}
Substituting the expression for $w(z)$ into Eq. \eqref{eq:fde_def} and using the series integral
\beq
\int_0^z \frac{{z'}^{\,n}}{1+z'}\,dz'
= (-1)^n \ln(1+z)
+ \sum_{k=0}^{n-1} (-1)^{k} \left(\frac{z^{n-k}}{n-k} \right)
= (-1)^n \ln(1+z)
+ \sum_{k=1}^{n} (-1)^{n-k} \left(\frac{z^k}{k} \right)
\eeq
one can calculate
\begin{align}
 f_{\de}(z)&=\exp \left\{ 3 \left[ \ln \left( 1+z \right) \left( 1+w_0 + \sum_{n=1} ^{\infty}(-1)^n \frac{w_n}{n!} \right) +  \sum_{n=1} ^{\infty} \frac{w_n}{n!} \left(\sum_{k=1}^n (-1)^{n-k} \frac{z^k}{k} \right)  \right]  \right\}  \\
 &= \left(1+z \right)^{3\left(1+w_0 \right)} \exp \left\{ 3 \left[ \ln \left( 1+z \right) \left(\sum_{n=1} ^{\infty}(-1)^n \frac{w_n}{n!} \right) + \sum_{n=1} ^{\infty} \frac{w_n}{n!} \left( \sum_{k=1}^n (-1)^{n-k} \frac{z^k}{k} \right)  \right]  \right\}
\end{align}
This is a generic expression valid for any smooth $w(z)$ at any $z$.

In the domain $z<1$, where the Taylor expansion of $\ln(1+z) = \sum_{m=1}^{\infty} (-1)^{m+1} \left(z^m / m \right)$ converges, we can rearrange the sums:
\begin{align}
 f_{\de}(z)&=\left(1+z \right)^{3\left(1+w_0 \right)} \exp \left[ 3 S(z)\right]\;,
\end{align}
where
\begin{align}
S(z)&\equiv \sum_{n=1} ^{\infty} \frac{w_n}{n!} \left( \sum_{k=n+1}^{\infty} \left(-1 \right)^{n+k+1} ~\frac{z^k}{k} \right) = \sum_{k=2} ^{\infty} \frac{z^k}{k} \left(\sum_{n=1}^{k-1} \left(-1 \right)^{n+k+1} \frac{w_n}{n!} \right) \;.
\end{align}
In the second equality, we have swapped the order of the sums over the triangular region in the $\{n,k\}$ grid, which is justified since the double series is convergent for $z<1$. The series $S(z)$ is a power series starting at second order, i.e. its leading term is of $\o (z^2)$.

Expanding,
\begin{align}
    \frac{1}{(1+z)^3 -1}&=\frac{1}{3z} \sum_{n=0}^{\infty} (-1)^n \left( z+ z^2/3\right)^n=\frac{1}{3z} \left[ 1 -z +\frac{2}{3} z^2 -\frac{1}{3} z^3 +\frac{1}{9}z^4 + \o (z^5)\right]\\
    &=\sum_{n=0}^{\infty} \sum_{k=0}^{n} (-1)^n \binom{n}{k} 3^{-(k+1)} z^{\,n+k-1}
\end{align}
one can quickly calculate the leading order term in RHS of Eq. \eqref{eq:w_0_0} as,
\begin{align}
    \wzp(z)\equiv \frac{Om(z)-1}{1-\om0} &= \frac{f_{\text{DE}}(z)-1}{(1+z)^3-1} -1 \label{eq:w0_1}\\
    &= -1+ \frac{1}{3} \left[ \frac{1}{z} -1 +\o (z) \right] \times \left[ (1+z)^{3(1+w_0)} \exp \left\{3 S(z) \right\} -1 \right] \\
    &=-1+ \frac{1}{3} \left[ \frac{1}{z} -1 +\o (z) \right] \times 
    \left(\left[1+3(1+w_0)z + \o (z^2) \right] \left[1+\o (z^2) \right] -1\right) \\
    &=-1+\frac{1}{3}\left[3(1+w_0) \right] +\o (z)=w_0 +\o (z)\;.
\end{align}
This result holds for any smooth $w(z)$. Therefore, evaluating $\wzp(z)$ in the limit $z \to 0$ directly yields the value of $w_0$.  Note that:
\begin{enumerate}
    \item $\wzp(z) \approx w_0$ only in the low-redshift limit, $z \ll 1$, where higher-order terms remain negligible. In the opposite limit, $z \gg 1$, particularly during the matter-dominated era, $\wzp(z)\to -1$ generically for any cosmological model that undergoes a matter-dominated phase prior to late-time acceleration.
    
    \item However, for the special case of $\Lambda$CDM, $\wzp(z) = -1 = w_0$ at all $z$. Thus, $\wzp(z)$ also serves as a null test for the \lcdm model, analogous to the $Om$ diagnostic.
  
    \item $S(z)=\o (z^2)$, implies $w_n$ for $n\geq1$ never contributes to the leading order term that isolates $w_0$.
    \item More generally, the $n$-th derivative, $w_n$, contributes only to terms of order $z^k$ with $k \ge n$. Consequently, higher order derivatives of $w(z)$ affect only higher order terms, with minimal impact at low $z$.

\end{enumerate}

Considering higher-order terms one gets,
\begin{align}\label{eq:wp_final}
    \wzp(z) \equiv \frac{Om(z)-1}{1-\om0} 
    &= w_0 + \left[\frac{3}{2}w_0(1+w_0) + \frac{w_1}{2}\right] z + \left[\frac{3}{2}w_0^2(1+w_0) + w_1 \left( \frac{3}{2}w_0 + \frac{2}{3}\right) + \frac{w_2}{6} \right] z^2 + \o (z^3)
\end{align}

It is important to emphasize that $\wzp(z)$ is not a reconstruction of $w(z)$ at redshift $z$. Rather, it provides an estimate of the present-day value $w_0$ using cosmological information available at finite redshift, since one cannot obtain direct cosmological constraints at exactly $z=0$.

\subsection{The novelty of $w_0$-probe}
In standard approaches, recovering the dark energy EoS $w(z)$ typically
involves differentiating the reconstructed $h(z)$, as in
Eq.~\eqref{eq:weff}. This is undesirable, since even small fluctuations in
$h(z)$ inferred from discrete data points can be strongly amplified, leading to
noisy and unstable estimates of $w(z)$.

The $w_0$-probe provides an alternative route, allowing a direct determination of the present-day EoS, $w_0$, from low-redshift reconstructions of $h(z)$ without requiring differentiation.

\section{Demonstration -- Data and Methodology}

In this section we demonstrate how the $w_0$-probe can be applied to real data. 
To obtain a model-independent estimate of $Om(z)$, and subsequently 
$\wzp(z)$, from the datasets described below, we first reconstruct 
the expansion history, $h(z)$, in a model-independent 
manner using Gaussian processes (GP).

\subsection{Datasets}

We use three observational datasets to reconstruct $h(z)$: 
\begin{itemize}
\item \textbf{Type Ia supernovae (SNe Ia):} We use the Union3 supernova dataset \citep{Rubin:2023jdq}.

\item \textbf{Baryon acoustic oscillations (BAO):} We use the DESI DR2 BAO measurements \citep{desi_dr2_cosmology}, which provide constraints on $D_M(z)/r_d$ and $D_H(z)/r_d$ in seven redshift bins
using five tracers. Here $D_M(z) = (1+z)D_A(z)$ and $D_H(z) = c/H(z)$ are the transverse comoving distance and the Hubble distance along the
line-of-sight. $r_d$ is the comoving sound horizon at
the baryon drag epoch, which serves as the standard ruler for BAO
measurements.

\item \textbf{Cosmic microwave background (CMB):}
We incorporate Planck constraints in a form suitable for
model-independent GP reconstructions
\citep{planck2018}. Specifically, we take the \lcdm fit to the Planck CMB data and compute the corresponding posterior
distributions for $D_H(z_*)$ and $D_A(z_*)$, where
$z_* \approx 1100$ is the redshift of the last-scattering surface.
These constraints on $D_H(z_*)$ and $D_A(z_*)$ are then used
as likelihoods when reconstructing $h(z)$. Even though this approach uses a $\Lambda$CDM fit, the assumption
is made at high redshift and does not impose a strong prior on the
late-time dynamics, allowing it to be used for dark-energy tests.
See \citet{late-kaplinghat1} for further details.

\end{itemize}

\subsection{Reconstructing expansion history using Gaussian process regression}
\label{sec:GP}

As stated above, in this work we reconstruct the expansion history, $h(z)$, using the Gaussian process (GP) ``prior approach'' following
\citet{late-kaplinghat1,Keeley:2020aym,Hwang:2022hla}.
This approach allows for a model-independent reconstruction
of the expansion history without assuming a specific
parametric form for $h(z)$. Specifically, we generate prior samples of $h(z)$ using a squared-exponential kernel,
\begin{equation} \label{eq:GP1}
\langle \gamma(s_1)\gamma(s_2) \rangle =
\sigma_f^2 \exp\!\left[-\frac{(s_1 - s_2)^2}{2\ell^2}\right],
\end{equation}
where $s(z) = \log(1+z)/\log(1+z_*)$ and $\gamma(z) = \log[h(z)/h_{\rm fid}(z)]$. We adopt a family of fiducial ``mean functions'', $h_{\rm fid} (z)$, rather than relying on a single choice, since the mean function can significantly influence GP reconstructions of the expansion history \citep{Hwang:2022hla}. These mean functions are constructed from CPL models with different $(w_0,w_a)$ values, interpolating between $\Lambda$CDM, $(w_0,w_a)=(-1,0)$, and the best-fit CPL model to the datasets described above, $(w_0,w_a)\approx(-0.61,-1.53)$. We select 25 anchor points along this line and interpolate the corresponding best-fit $(\om0,H_0)$ values to determine these parameters for each sampled $(w_0,w_a)$ pair. The hyperparameters $\sigma_f$ and $\ell$ control the overall amplitude of deviations and the correlation scale of the reconstructed function, respectively. We vary $\sigma_f$ over $[10^{-3},10^{-1}]$ and $\ell$ over $[10^{-2},1]$, with uniform priors on $(\log\sigma_f,\log\ell)$.

The reconstructed expansion history for a given mean function is then written as
\begin{equation}
h(z) = h_{\rm fid}(z)\exp[\gamma(z)] .
\end{equation}
The covariance function in Eq. \eqref{eq:GP1} defines a probability distribution
over functions $\gamma(z)$ from which we draw samples,
thereby generating a family of candidate expansion histories that
serve as GP priors. We combine the resulting realizations across the family of mean functions and compute their likelihoods against the observational datasets described above, allowing us to identify the range of expansion histories consistent with the data.

Once the likelihoods for the ensemble of reconstructed $h(z)$
functions are computed, there are several possible ways to
proceed in quantifying the constraints from these reconstructions.
In this work we consider two approaches, briefly outlined below:

\begin{itemize}
\item \textbf{Gaussian Process posterior samples:}
Each reconstructed $h(z)$ realization is weighted by its likelihood. The posterior distributions of $h(z)$, or of derived quantities such as the $\wzp(z)$, are then used to construct the corresponding credible regions (e.g., $68\%$ and $95\%$ intervals).

\item \textbf{$\chi^2$-limited samples:}
We instead display only the subset of reconstructed $h(z)$ realizations that provide `acceptable' fits to the data, defined as those having likelihoods exceeding the $95\%$ confidence level (C.L.) threshold (i.e. the $2\sigma$ limit) of the CPL \citep{CPL1,CPL2} fit to the same datasets.

\end{itemize}

We describe these methods in greater detail in the next section,
together with the results obtained using each approach.


\section{Results}
\subsection{Constraints on $h(z)$ and $Om(z)$ from the Gaussian Process posterior samples}

\begin{figure}
 \centering
\includegraphics[width=0.4855\linewidth]{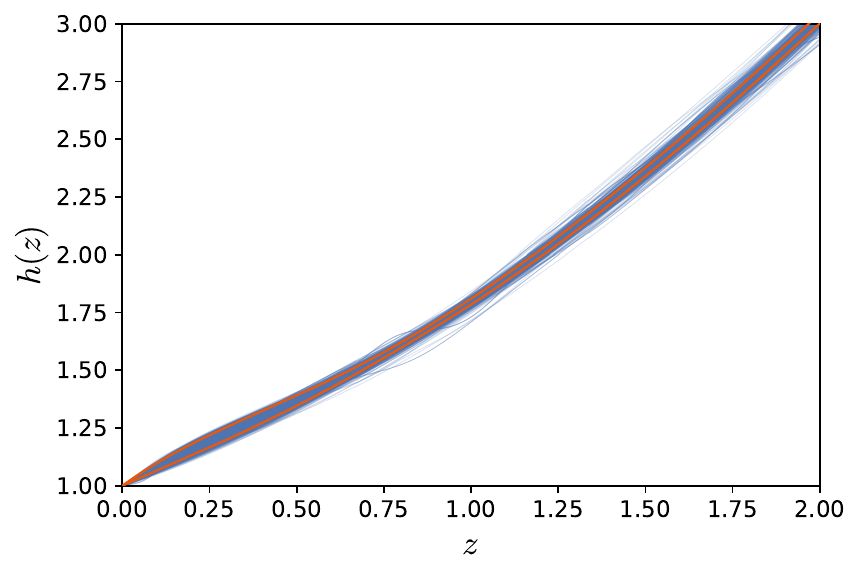}
\includegraphics[width=0.4855\linewidth]{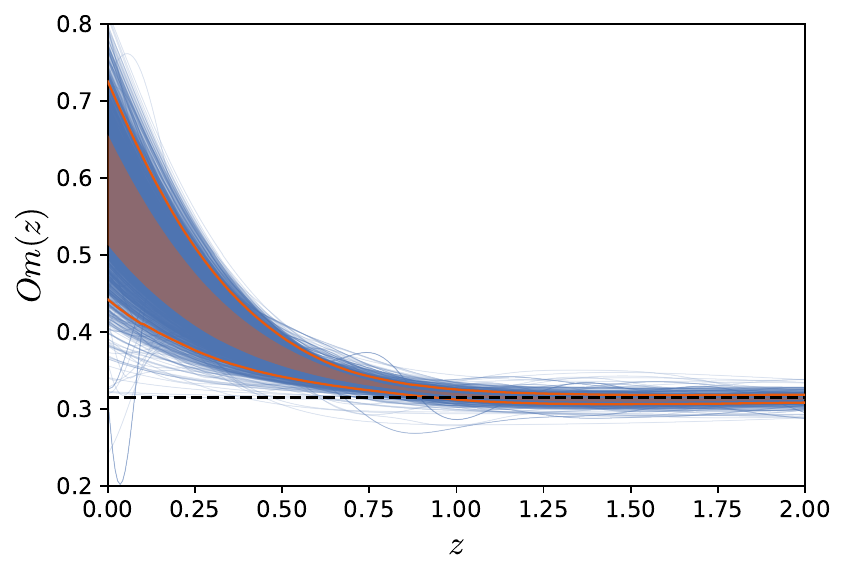}
\caption{Samples of $h(z)\equiv H(z)/H_0$ and $Om(z)$, drawn from the Gaussian process posterior distribution, are shown in the left and right panels respectively. The orange lines in both panels enclose the $95\%$ credible region, while the orange shading, shown only in the right panel, indicates the $68\%$ level.
}
\label{fig:hz_Omz}
\end{figure}

\begin{figure}
 \centering
\includegraphics[width=0.5\linewidth]{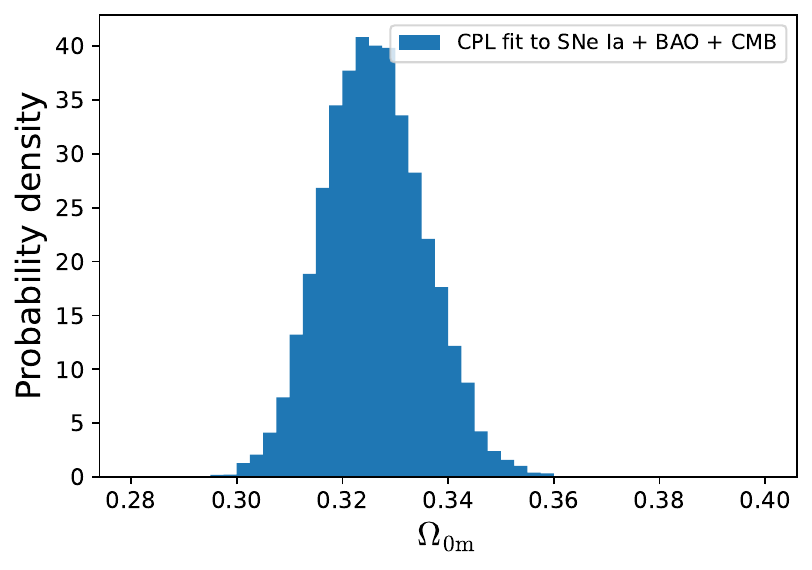}
\caption{Posterior distribution of $\om0$ from fitting the CPL parameterization to SNe~Ia+BAO+CMB data. We marginalize over this conservative $\om0$ distribution to estimate $\wzp(z)$ as shown in the left panel of Figs. \ref{fig:wp_hpost} and \ref{fig:wp_chi2_limited}.  
}
\label{fig:Om0_dist}
\end{figure}

The most straightforward approach follows the standard Bayesian
procedure. Each generated $h(z)$ realization is weighted by its
likelihood, and at every redshift we construct the posterior
distribution of $h(z)$, or of any derived quantity such as
$Om(z)$ or the $\wzp(z)$. From these distributions we obtain
the corresponding confidence levels (CLs). Connecting the CL
intervals across redshift produces the familiar band plots that
are commonly used to visualize reconstructed expansion histories.

The left panel of Fig.~\ref{fig:hz_Omz} shows samples of $h(z)$ drawn from the
Gaussian process posterior distribution, while the right panel displays the
corresponding $\omz$ samples. In both panels, the orange lines enclose the
$95\%$ credible region, while the orange shading in the right panel indicates
the $68\%$ level. In the right panel, the horizontal dashed black line
indicates the Planck--\lcdm value $\om0 = 0.315$ \citep{planck2018}. 

The reconstruction shows that the Planck–$\Lambda$CDM expectation lies outside the $95\%$ credible region of $Om(z)$ at low redshifts. 
In particular, $Om(z)$ exhibits a rising trend relative to
$\om0 = 0.315$ at $z \lesssim 0.7$. In the limit $z\to0$, only $\simeq0.13\%$ of the posterior probability
lies below $\om0=0.315$, corresponding to an exclusion of \lcdm at a one-sided
Gaussian-equivalent significance of $\simeq3\sigma$. When compared with the
theoretical $Om(z)$ curves for different cosmological models
(see Fig.~\ref{fig:omz_models}), such behavior is qualitatively
consistent with models having $w_0 > -1$. However, this feature
is not uniquely diagnostic and alternative scenarios remain
possible.

\subsection{Treatment of $\om0$ in $\wzp(z)$ calculation}
Inferring $\wzp(z)$ from $Om(z)$ additionally requires the present matter density, $\om0$, whose true model-independent value is not directly measured. This also spoils the traditional way described in Eq. \eqref{eq:weff}. Nevertheless, an incorrect choice of $\om0$ by an amount
$\delta\om0$ simply rescales the $w_0$-probe as
\begin{equation}
\wzp(z) \;\rightarrow\; \wzp(z) \left(1+\frac{\delta\om0}{1-\om0 - \delta\om0}\right) \approx \wzp(z) \left(1+\frac{\delta\om0}{1-\om0}\right),
\end{equation}
provided $\delta\om0\ll(1-\om0)$. For instance, a scatter of $\delta\om0\sim0.01$ around $\om0\approx0.32$,
typical of the $1\sigma$ uncertainty in the CPL fit (Fig.~\ref{fig:Om0_dist}),
rescales the $\wzp$ by $\lesssim1.5\%$. Thus, moderate uncertainties in $\om0$ primarily induce a multiplicative rescaling of $\wzp(z)$, without altering its redshift dependence. When marginalizing over a reasonable unbiased posterior for $\om0$, with $\langle\delta\om0\rangle=0$, the resulting estimate of $\wzp(z)$ remains unbiased, with the uncertainty in $\om0$ contributing only to an increased scatter among samples.

In practice, we adopt two complementary approaches:
\begin{itemize}
    \item \textbf{CPL-$\om0$ marginalization}: We obtain a conservative sample of $\om0$ by fitting the flexible CPL model to the combined SNe Ia+BAO+CMB datasets. The resulting distribution is shown in Fig.~\ref{fig:Om0_dist}. At a given redshift $z$, samples of $\wzp(z)$ are then generated by evaluating Eq.~\eqref{eq:w0probe} for each reconstructed $Om(z)$ and for each value of $\om0$ drawn from this sample.

    \item \textbf{Radiation subtraction}: 
Since the CMB temperature is measured with high precision, 
the quantity $\omegr0 H_0^2$
is accurately determined once the standard assumptions about neutrinos are adopted, namely $N_{\rm eff}=3.046$ with one minimally massive neutrino.
At sufficiently high redshift, $z \sim \mathcal{O}(10^3)$,
the contribution of dark energy to $H(z)$ is negligible, so that
$\om0 H_0^2 (1+z)^3 \simeq H^2(z) - \omegr0 H_0^2 (1+z)^4$.
This relation therefore allows us to infer $\om0 H_0^2$
directly from the reconstructed $H(z)$ at high redshift. The
method assumes that no additional energy component contributes
significantly to $H(z)$ at early times, and relies on the fact
that the Gaussian process regression described above allows the
reconstruction of $H(z)$ to be extended up to $z \sim \mathcal{O}(10^3)$.
\end{itemize}

\subsection{$\wzp(z)$ from the Gaussian Process posterior samples}\label{sec:om0_marg}

\begin{figure}
 \centering
\includegraphics[width=0.4855\linewidth]{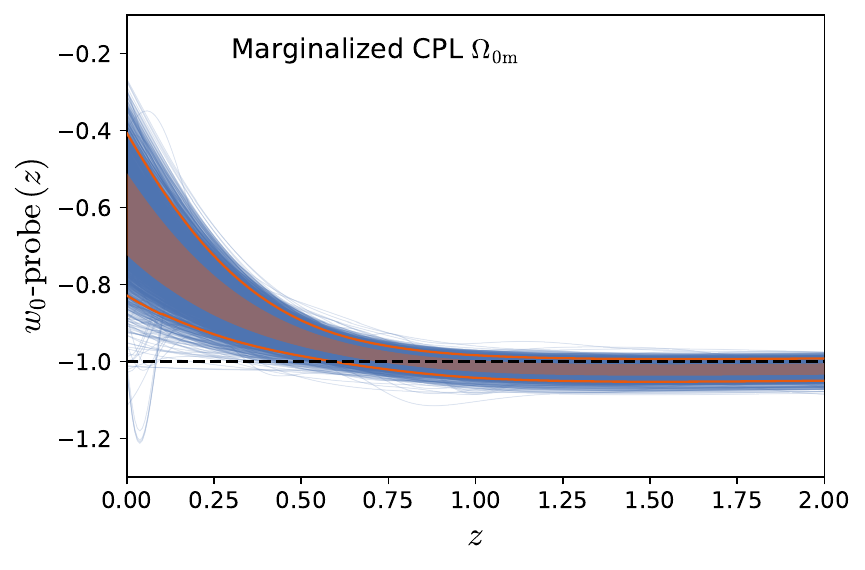}
\includegraphics[width=0.4855\linewidth]{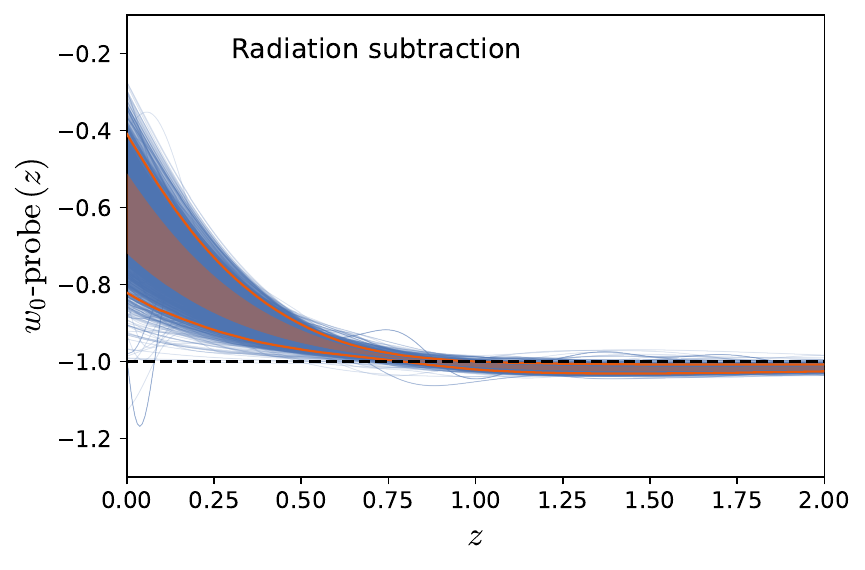}
\caption{$w_0$-probe estimates obtained using two treatments of $\om0$: CPL $\om0$ marginalization (left) and radiation subtraction (right). In both panels, the orange shaded regions and orange lines indicate the $68\%$ and $95\%$ credible regions, respectively. The two approaches yield consistent results, with the CPL-marginalized case exhibiting larger scatter due to $\om0$ uncertainty. In both cases, $w_0=-1$ ($\Lambda$CDM) lies outside the $95\%$ credible region, and the reconstruction favors $w_0 \simeq -0.62_{-0.10}^{+0.11}$ at $68\%$ C.L. Note that $\wzp(z)$ should not be interpreted as a reconstruction of $w(z)$; it estimates the present-day value $w_0$ using data at finite redshift.}

\label{fig:wp_hpost}
\end{figure}

The $w_0$-probe results for the two approaches are shown in Fig.~\ref{fig:wp_hpost}. In the left panel, we marginalize over the CPL-fit $\om0$ posterior while evaluating $\wzp(z)$ using Eq.~\eqref{eq:w0probe}. In the right panel, we apply the radiation-subtraction method described above. In both panels, the orange shading and orange lines indicate the $68\%$ and $95\%$ credible regions,
respectively. The two approaches yield mutually consistent results, with the radiation-subtraction method providing slightly tighter constraints since it does not involve marginalization over a broad $\om0$ distribution.

In both approaches, the $\wzp$ remains close to the \lcdm value,
$\wzp(z)\simeq -1$, at sufficiently high redshifts, $z\gtrsim0.7$, consistent
with the generic matter-dominated limit of the probe given by
Eq.~\eqref{eq:wzp_mat_dom}, and therefore not necessarily indicative of
$\Lambda$CDM itself. At lower redshifts, however, $\wzp(z)$ exhibits a clear
evolving, non-flat behavior, signaling an inconsistency of the datasets with
the \lcdm model, in agreement with the conclusions drawn from $\omz$. Moreover,
in the limit $z\ll1$, $\wzp(z)$ directly traces the true $w_0$. Evaluating $\wzp(z)$ at $z\to0$ in both panels yields
$w_0=-0.62^{+0.11}_{-0.10}$ at $68\%$ C.L., with a $95\%$ credible interval of
$[-0.83,-0.41]$, significantly higher than the \lcdm value. Only $\simeq0.14\%$
of the posterior probability lies below $w_0=-1$, in both treatments of $\om0$,
corresponding to an exclusion of \lcdm at a Gaussian-equivalent significance
of $\simeq3\sigma$. Note that this estimate of $w_0$ is consistent with the
direct CPL fit to the same combination of datasets
\citep{desi_dr2_cosmology}, although here it is obtained without assuming a
parametric form for $w(z)$. The exclusion of \lcdm at this level is broadly
consistent with several recent studies combining DESI data with Planck CMB
measurements and different SNe~Ia compilations
\citep{desi_dr2_cosmology, Rodrigo2024, lodha2025}.

\subsection{Considering $\chi^2$-limited samples}
\begin{figure}
 \centering
\includegraphics[width=0.4855\linewidth]{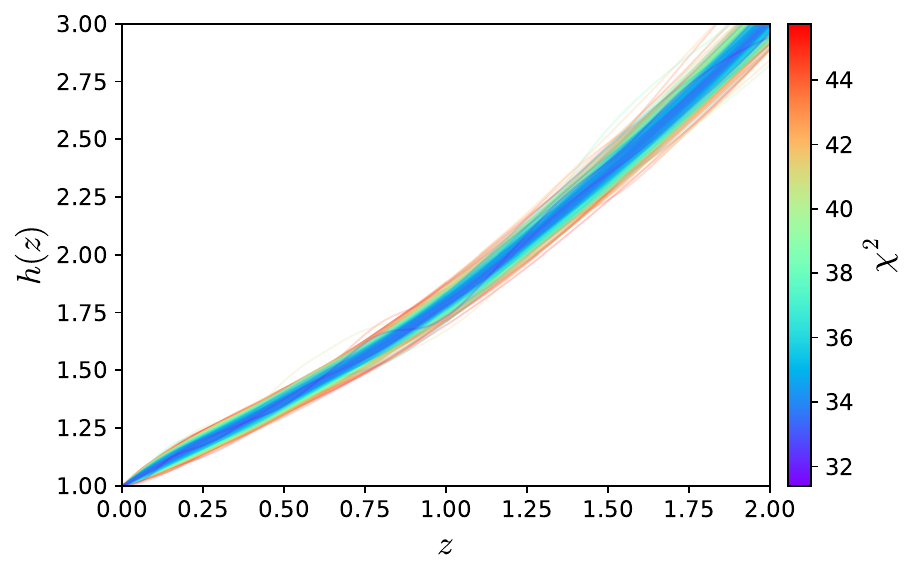}
\includegraphics[width=0.4855\linewidth]{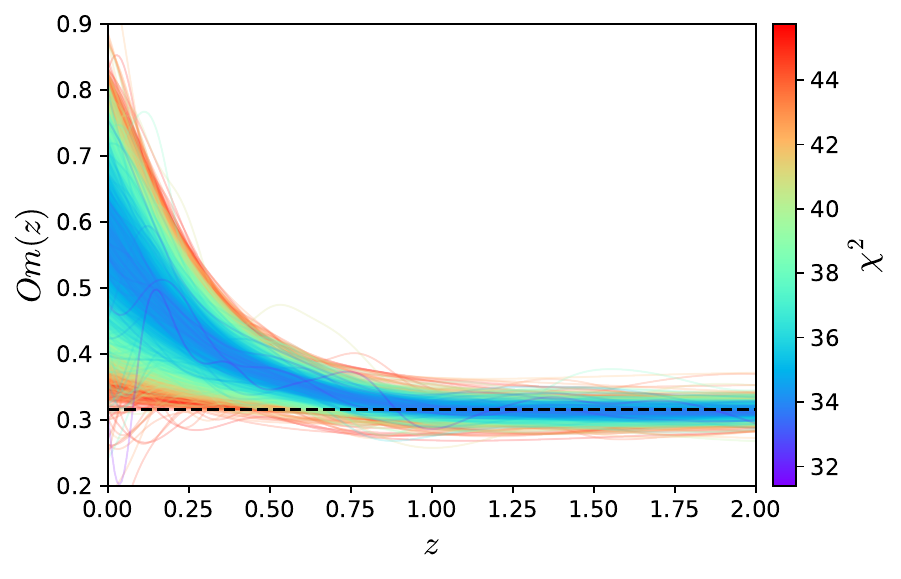}
\caption{$\chi^2$-limited samples of $h(z)$ (left panel), selected to have likelihoods exceeding the $95\%$ confidence threshold of the CPL fit, and the corresponding $Om(z)$ curves (right panel). Curves are color-coded by their $\chi^2$ values (see colorbars). The higher-likelihood $Om(z)$ samples consistently exclude the \lcdm value $\om0=0.315$ for $z\lesssim0.5$, approximately since the matter–dark energy equality epoch.}
\label{fig:hz_Omz_chi2_limited}
\end{figure}

\begin{figure}
 \centering
\includegraphics[width=0.4855\linewidth]{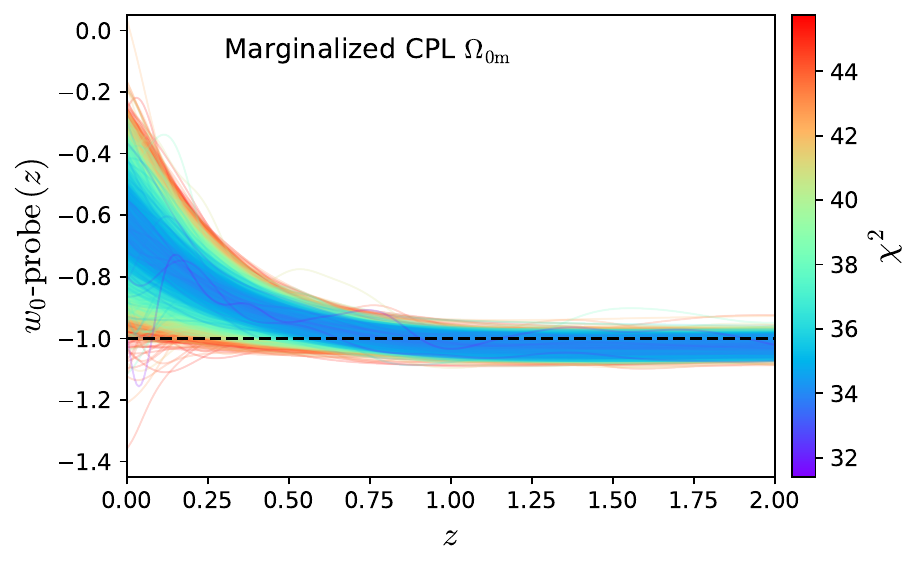}
\includegraphics[width=0.4855\linewidth]{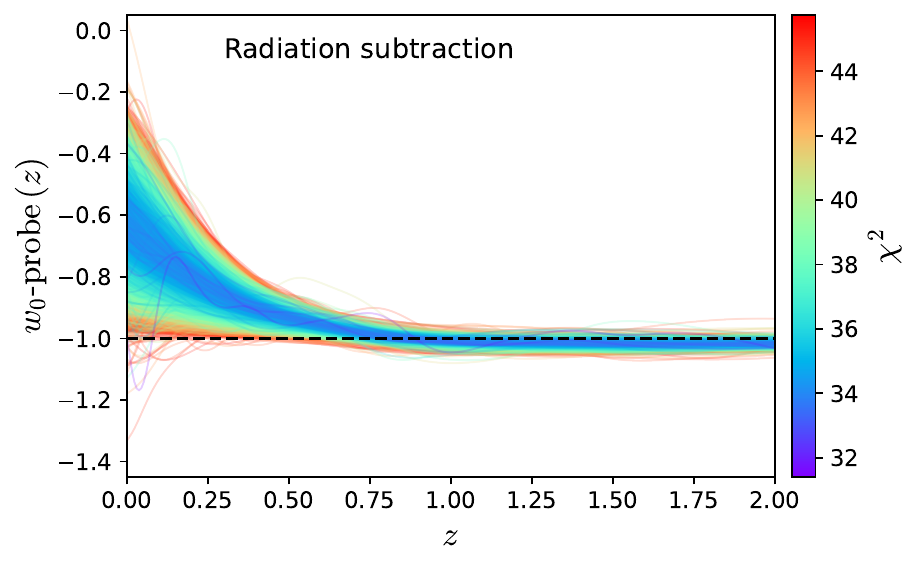}
\caption{$w_0$-probe estimates obtained from the $\chi^2$-limited samples shown in Fig.~\ref{fig:hz_Omz_chi2_limited}. Results from two treatments of $\om0$, CPL $\om0$ marginalization (left panel) and radiation subtraction (right panel), are shown. Curves are color-coded by their $\chi^2$ values. The two approaches yield consistent results, with the CPL-marginalized case exhibiting larger scatter due to $\om0$ uncertainty. At low redshifts, $z \lesssim 0.5$, the highest-likelihood samples exclude $w_0=-1$, and hence the \lcdm model. Note that $\wzp(z)$ should not be interpreted as a reconstruction of $w(z)$; it estimates the present-day value $w_0$ using data at finite redshift.}
\label{fig:wp_chi2_limited}
\end{figure}

The above results are based on the GP posterior samples of $h(z)$. To assess the robustness of our conclusions, we perform an independent analysis using $\chi^2$-limited samples of $h(z)$. Specifically, we retain all GP-reconstructed $h(z)$ realizations satisfying $\chi^2 < \chi^2_{\rm th}$, where the threshold is set by the $95\%$ confidence level of the CPL fit to the datasets. Unlike the posterior samples, these samples are not weighted by $\propto \exp(-\chi^2/2)$; instead, all realizations passing the threshold are assigned equal weight. This ensures that only reconstructions with likelihoods comparable to or better than the $95\%$ CPL quantile are retained, thereby relaxing any potential GP prior-induced constraints.

The resulting $\chi^2$-limited $h(z)$ samples are shown in the left panel of Fig.~\ref{fig:hz_Omz_chi2_limited}, while the corresponding $Om(z)$ realizations are displayed in the right panel. The color scale indicates the $\chi^2$ value of each sample, with blue-violet curves representing the highest-likelihood reconstructions. 

The $w_0$-probe inferred from these $\chi^2$-limited samples is shown in Fig.~\ref{fig:wp_chi2_limited}. The left and right panels correspond to the two treatments of $\om0$, namely CPL $\om0$ marginalization (left) and radiation subtraction (right), with curves again color-coded by their $\chi^2$ values. As expected, the CPL $\om0$ marginalization produces a slightly broader region compared to the radiation-subtraction method, although the two approaches remain mutually consistent, mirroring the behavior seen in Fig.~\ref{fig:wp_hpost}.

We now focus on the realizations best supported by the data, which we define
as those within a likelihood ratio of 100 of the best-fitting reconstruction,
i.e.\ $\Delta\chi^2 \equiv \chi^2-\chi^2_{\rm min} \le 2\ln 100 \simeq 9.2$.
These high-likelihood realizations, which constitute about $60\%$ of the
$\chi^2$-limited samples, start to exclude $w_0=-1$ below $z\simeq0.5$. In the limit $z\to0$, more than $99.9\%$ of them lie above $w_0=-1$, and the
central $95\%$ of their $\wzp$ values lie in $w_0\in(-0.85,\,-0.40)$ for both
treatments of $\om0$. Notably, this range is in close agreement with the $95\%$ credible
interval $[-0.83,\,-0.41]$ obtained from the GP posterior samples in
Sec.~\ref{sec:om0_marg}, even though the two analyses weight the
reconstructions differently. Since the $\chi^2$-limited samples are not
weighted by their likelihood, this range characterizes the spread of the
high-likelihood reconstructions rather than a formal statistical confidence
interval.

\section{Conclusion and Discussions}
Determining the equation of state (EoS) of dark energy is of fundamental importance, particularly in light of recent analyses of DESI data that hint at possible dynamical behavior. However, commonly used parametrizations such as the CPL model are phenomenological and are unlikely to capture the true underlying physics of dark energy, raising the possibility that certain inferred features, for example phantom crossing, may arise from the parametrization itself rather than the data. Model-independent approaches, together with diagnostic tools, such as the $Om$ diagnostic, are therefore essential.

Traditionally, the dark energy EoS is reconstructed by differentiating the expansion history inferred from discrete data, a procedure that is inherently noisy and unstable. In this work, we introduce a new diagnostic, the $w_0$-probe which is constructed from the $Om$ diagnostic very simply via (\ref{eq:w0probe}). In the low redshift limit when $z\to 0$, the $w_0$-probe reduces to the dark energy EoS $w_0$. Therefore a precise knowledge of the $Om$ diagnostic at low redshifts allows one to reconstruct $w_0$. Since $Om(z)$ depends only upon the expansion history, $h(z)$, so does the $w_0$-probe. This enables the $w_0$-probe to directly determine the present-day EoS, $w_0$, solely from observations of $h(z)$ without any additional differentiation. While $Om(z)$ itself provides only a null test of $\Lambda$CDM, the $w_0$-probe not only retains this capability but also allows direct estimation of $w_0$. We have shown rigorously that this result holds for any underlying smooth $w(z)$.

Moreover, unlike conventional reconstructions based on Eq.~\eqref{eq:weff} the $w_0$-probe is affected by uncertainties in $\om0$ only through a multiplicative rescaling. Consequently, marginalizing over a reasonable unbiased posterior for $\om0$ yields an unbiased estimate of the true $w_0$, with $\om0$ uncertainties contributing solely to the overall scatter. We explored two complementary treatments of $\om0$: CPL $\om0$ marginalization and radiation subtraction.

We demonstrated the application of this method using GP-reconstructed expansion histories from SNe Ia+BAO+CMB data. The reconstructed $Om(z)$ consistently excludes the Planck–\lcdm value $\om0=0.315$ for $z\lesssim0.5$, approximately coincident with the epoch when dark energy becomes dynamically significant relative to matter. The $w_0$-probe naturally captures this deviation, as it inherits the same redshift dependence.

Using the GP posterior samples of $h(z)$, the $\wzp$ in the limit
$z\to0$ yields $w_0 \approx -0.62^{+0.11}_{-0.10}$ at $68\%$
C.L., with a $95\%$ credible interval of $[-0.83,-0.41]$, substantially higher
than the \lcdm value of $w_0=-1$. In this limit, only $\simeq0.13$--$0.14\%$ of
the posterior probability of both $Om$ and $\wzp$ lies on the \lcdm side,
corresponding to an exclusion of \lcdm at a Gaussian-equivalent significance
of $\simeq3\sigma$. This inferred value is consistent with the direct CPL fit
to the same combination of datasets \citep{desi_dr2_cosmology}. The exclusion
of \lcdm at this level is also consistent with several recent analyses
combining DESI data with Planck CMB measurements and different compilations of
SNe Ia data \citep{desi_dr2_cosmology, Rodrigo2024, lodha2025}. Importantly,
our estimate is obtained without assuming a parametric form for $w(z)$,
highlighting both the precision and the diagnostic power of the $\wzp$.

Since GP reconstructions may be influenced by prior choices, we additionally
examined $\chi^2$-limited $h(z)$ samples, retaining only realizations with
likelihoods exceeding the $95\%$ confidence threshold of the CPL fit. The
realizations best supported by the data, those with
$\Delta\chi^2 \le 2\ln 100 \simeq 9.2$ relative to the best-fitting
reconstruction, systematically exclude \lcdm in the limit $z\to0$: more than
$99.9\%$ of them lie above $w_0=-1$, and the central $95\%$ lie in
$w_0\in(-0.85,\,-0.40)$ for both treatments of $\om0$. This range, in close
agreement with the $95\%$ credible interval from the GP posterior samples,
provides a robust estimate of the present-day dark energy EoS, rather than a
formal statistical confidence interval.

\section*{Acknowledgment} 
SB is supported by the Deutsche Forschungsgemeinschaft (DFG, German Research Foundation) – SFB 1258 – 283604770. SB also acknowledges the support provided by the Alexander von Humboldt Foundation. VS thanks the Anusandhan National Research Foundation (ANRF), India, for the National Science Chair Professorship which provided
partial funding for this work.


\appendix

\section{\texorpdfstring{\wprobetitle}{w0-probe} for a few models}\label{app:w0probe_extra}

\begin{figure*}
 \centering
\includegraphics[width=0.325\textwidth]{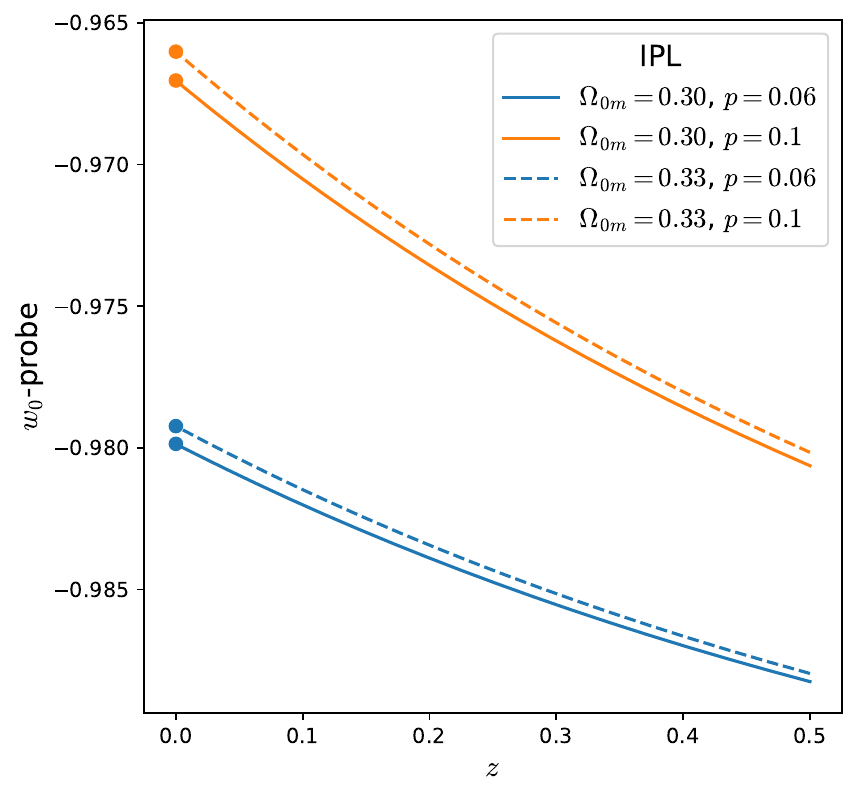}
\includegraphics[width=0.325\textwidth]{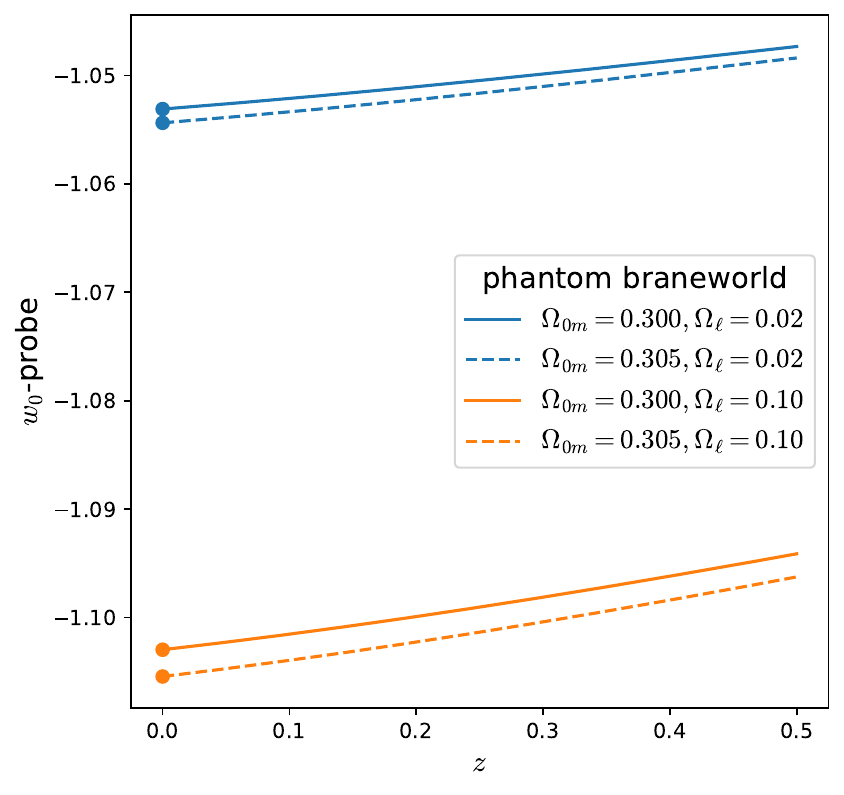}
\includegraphics[width=0.325\textwidth]{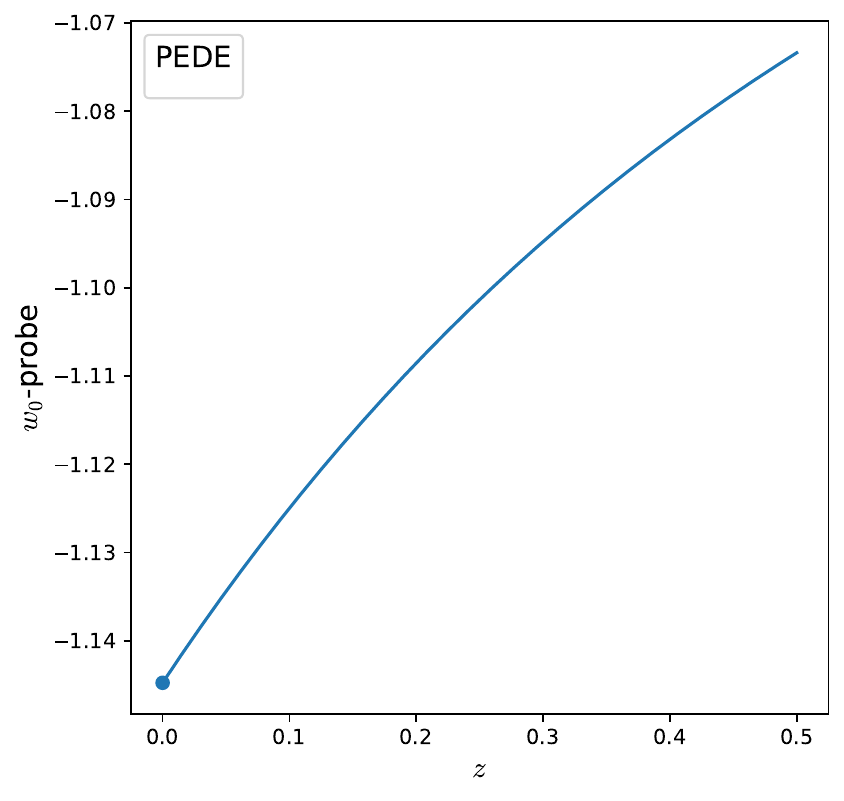}
\includegraphics[width=0.325\textwidth]{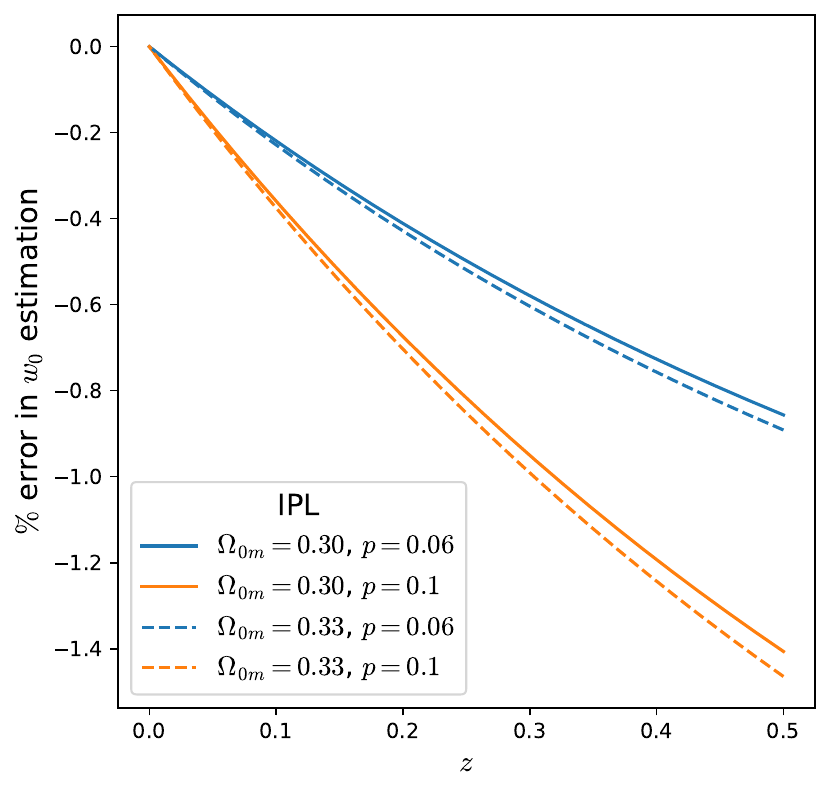}
\includegraphics[width=0.325\textwidth]{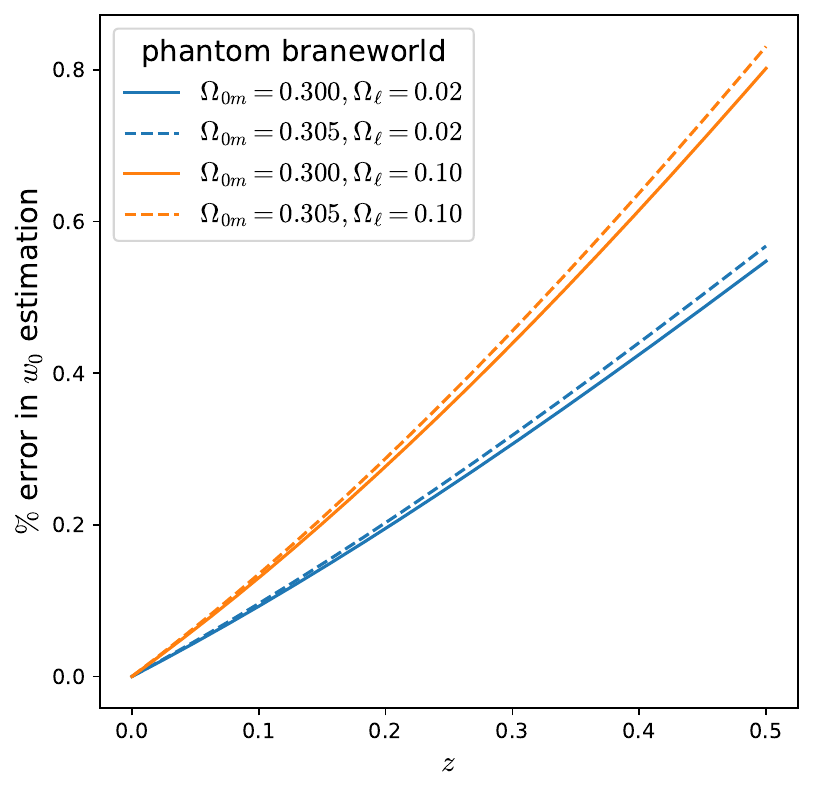}
\includegraphics[width=0.325\textwidth]{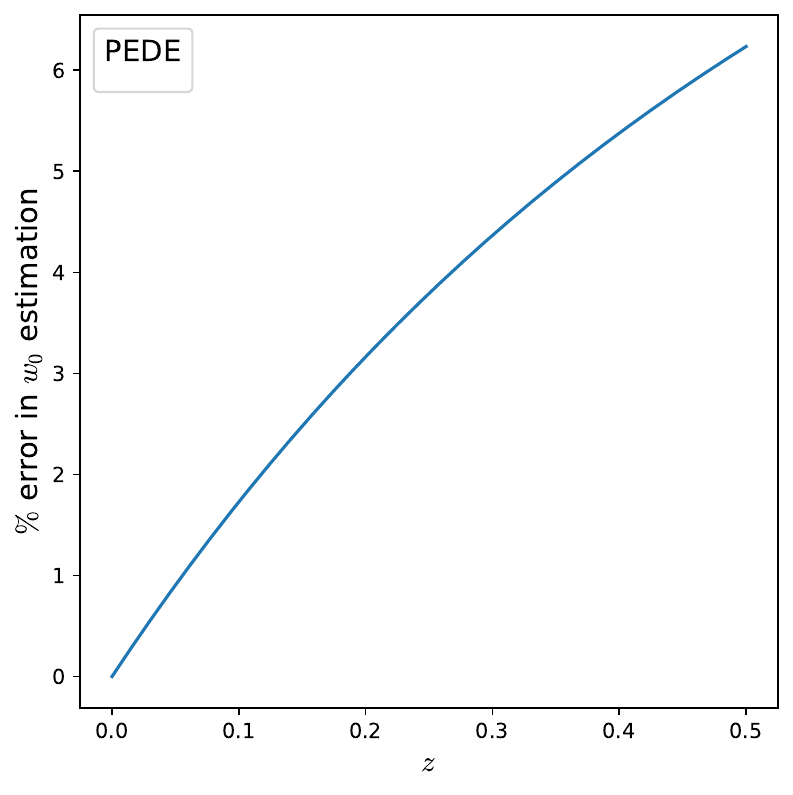}
\caption{{\bf Top panels} show the $w_0$-probe estimations using Eq. \eqref{eq:w0probe} for three different models -- inverse power-law (IPL) quintessence model, phantom braneworld and phenomenological emergent dark energy (PEDE) -- from left to right. In IPL and phantom braneworld models, the relevant parameter values are set to the limits that are consistent with the current observations. For PEDE, which does not have any extra degree of freedom as compared to $\Lambda$CDM,  $w_0$-probe is independent of any parameter such as $\om0$ since $w(z)$ is a fixed function of $z$. The true $w_0$ is marked by the filled circle on each plot, the curves show the $w_0$-probe estimations as a function of $z$. {\bf Bottom panels} show, for the same three models and parameter values, the percentage error in estimating $w_0$ with the $w_0$-probe at different redshifts, computed using Eq. \eqref{eq:Perror}. Naturally, when $z \to 0$, $w_0$-probe can recover the true $w_0$ very accurately, however, we find that the theoretical error remains below a few percent even at $z \sim 0.5$ for all these three models.}
\label{fig:wprobe_ibp}
\end{figure*}

We now consider several examples of evolving dark energy models: inverse power-law quintessence ($\phi$CDM) \citep{Ratra:1987rm, Bag:2017vjp}, the phantom braneworld model \citep{ss02,Bag:2021cqm}, and phenomenological emergent dark energy (PEDE) \citep{Li:2019yem}. In Fig. \ref{fig:wprobe_ibp}, we compare the $w_0$-probe results with the corresponding true values of $w_0$ for these models. The top panels show the left-hand side of Eq. \eqref{eq:w0probe} as a function of $z$ for different parameter choices, while the true $w_0$ values are indicated by filled circles. The bottom panels display the percentage error in the inferred $w_0$, computed using the left-hand side of Eq. \eqref{eq:w0probe}:

\beq\label{eq:Perror}
\text{Percentage error in } w_0 =
\frac{1}{|w_0|}
\left[\underbrace{\frac{Om(z)-1}{1-\om0}}_{\wzp(z)} - \ w_0 \right]\times100
\eeq

The deviation of the $w_0$-probe, $|\wzp(z)-w_0|$, may be interpreted as a theoretical or systematic error that depends on redshift and quantifies the intrinsic accuracy of the $w_0$-probe for a given underlying model.

It is evident from Figure \ref{fig:wprobe_ibp} that for these models, the $w_0$-probe, defined in Eq. \eqref{eq:w0probe}, measures the true present day EoS of dark energy ($w_0$) at very low redshifts, i.e., $\wzp(z) \to w_0$ as $z \to 0$, in all models, for all combinations of parameters. Nevertheless, $w_0$-probe estimates the true $w_0$ to within a few percent for $z \lesssim 0.5$, even for the most unfavorable parameter choices, provided the correct value of $\om0$ is adopted.

\begin{itemize}
\item Even for a constant $w(z)=w_0$, i.e. for $w_0$CDM model, Eq. \eqref{eq:wp_final} reduces to
\beq \label{eq:w0probe_w0}
\wzp(z) \equiv \frac{Om(z) - 1}{1 - \om0}=w_0+ z\left[\frac{3}{2} w_0 ( 1+ w_0 ) \right] + \o (z^2) \neq w_0\;,
\eeq
Thus, $\wzp(z)$ is generally not constant, even when $w$ itself is constant. However, the coefficient of the leading-order correction term, $w_0(1+w_0)$, vanishes as $w_0 \to -1$ or $w_0 \to 0$. In fact, these two values represent special fixed points for which $\wzp(z)=w_0$ exactly at all redshifts, as explained by Eq.~\eqref{eq:wzp_fixed_pts}.

\item For the CPL parametrization \citep{CPL1,CPL2},
\beq
w(z) = w_0 + w_1\frac{z}{1+z}~,~ \left.\frac{dw}{dz}\right|_{z=0}=w_1 \;,
\label{eq:cpl}
\eeq
one obtains from Eq.~\eqref{eq:wp_final}
\beq \label{eq:w0probe_cpl}
\wzp(z)=w_0+ z\left[\frac{3}{2} w_0 ( 1+ w_0) + \frac{1}{2} w_1 \right] + \o (z^2)\;.
\eeq

Thus, for the CPL parametrization, the leading contribution to the theoretical error in the $w_0$-probe (i.e. the mismatch between the $\wzp(z)$ and the true value of $w_0$ as given by Eq. \eqref{eq:Perror}) arises from the linear term 
$z\left[\frac{3}{2} w_0 (1+w_0) + \frac{1}{2} w_1 \right]$. 

Observationally, $w_0$ is already known to be close to $-1$, so the contribution from the $w_0(1+w_0)$ term is typically small. In contrast, $w_1$ is still poorly constrained by current datasets. Consequently, a large $w_1$ can induce a significant bias in the $w_0$-probe. The largest error occurs when the signs of $w_1$ and $w_0(1+w_0)$ are the same, which happens for two cases: (i) $w_0>-1$ with a large negative $w_1<0$, and (ii) a phantom dark energy today ($w_0<-1$) with a large positive $w_1>0$. Both of these combinations necessarily imply a crossing of the phantom divide in the past. In the remaining two sign combinations the contributions partially cancel, resulting in a much smaller theoretical error.
\end{itemize}

\bibliographystyle{aasjournal}
\bibliography{ref}

\end{document}